\input{aipcheck}

\documentclass[
    ,final            
  ]
  {aipproc}

\layoutstyle{8x11single}

\def\apj{ApJ.}

\def\mnras{MNRAS}

\begin{document}

\title{Variability in the Prompt Emission of Swift-BAT Gamma-Ray Bursts}



\author{T. N. Ukwatta}{
  address={The George Washington University, Washington, D.C. 20052}
  ,altaddress={NASA Goddard Space Flight Center, Greenbelt, MD 20771}
}

\author{K. S. Dhuga}{
  address={The George Washington University, Washington, D.C. 20052}
}

\author{W. C. Parke}{
  address={The George Washington University, Washington, D.C. 20052}
}

\author{T. Sakamoto}{
  address={NASA Goddard Space Flight Center, Greenbelt, MD 20771}
  ,altaddress={The University of Maryland, Baltimore County, Baltimore, MD 21250}
}

\author{C. B. Markwardt}{
  address={NASA Goddard Space Flight Center, Greenbelt, MD 20771}
}

\author{ S. D. Barthelmy}{
  address={NASA Goddard Space Flight Center, Greenbelt, MD 20771}
}

\author{ D. F. Cioffi}{
  address={The George Washington University, Washington, D.C. 20052}
}

\author{ A. Eskandarian}{
  address={The George Washington University, Washington, D.C. 20052}
}

\author{ N. Gehrels}{
  address={NASA Goddard Space Flight Center, Greenbelt, MD 20771}
}

\author{ L. Maximon}{
  address={The George Washington University, Washington, D.C. 20052}
}

\author{D. C. Morris}{
  address={The George Washington University, Washington, D.C. 20052}
  ,altaddress={NASA Goddard Space Flight Center, Greenbelt, MD 20771}
}


\begin{abstract}
We present the results of our study of the variability time scales
of a sample of 27 long Swift Gamma-Ray Bursts (GRBs) with known
redshifts. The variability time scale can help our understanding
of fundamental GRB parameters such as the initial bulk Lorentz
factor and the characteristic size associated with the emission
region. Fast Fourier Transform (FFT) techniques were used to
extract a noise threshold crossing frequency, which we associate
with a variability time scale. The threshold frequency appears to
show a correlation with the peak isotropic luminosity of GRBs.
\end{abstract}

\maketitle


\section{Introduction}

The time variability in Gamma-ray Bursts (GRBs) is crucial to our
understanding of the characteristic size associated with GRBs.
\\ \\
In 2000, Fenimore and Ramirez-Ruiz first proposed a correlation
between variability of GRBs and peak isotropic
luminosity~\cite{Fenimore2000}. Since then a number of authors
have provided further support for this correlation
\cite{Reichart2001,Guidorzi2005a, Guidorzi2005b, Guidorzi2006,
Li2006, Rizzuto2007}. However, these authors used a variety of
definitions for variability with various parameters and smoothing
methods. The lack of a universally accepted definition for
variability is a major short coming and poses problems in
comparing and evaluating the results of previous studies.
\\ \\
In our analysis we have used Fourier analysis techniques to probe
various frequencies and their strength in GRB light curves. In
this paper we associate a threshold frequency, which is the
frequency where the signal crosses the noise level, as a potential
variability indicator.
\\ \\
Due its fast slewing capability, the $Swift$ Gamma-Ray Burst
mission~\cite{gehrels2004} has enabled more redshift measurements
of GRBs than before. With its highly sensitive primary instrument,
the Burst Alert Telescope (BAT)~\citep{barthelmy2005}, $Swift$
provides high quality GRB data on which a Fourier analysis can be
performed and a variability time scale extracted. Our sample
contains 27 $Swift$ BAT GRBs with good spectral measurements. We
extracted the threshold frequency for all the GRBs and show this
parameter is correlated with the isotropic peak luminosity.

\section{Methodology}
\begin{figure}[h]
\includegraphics[angle=0, width=0.6\textwidth]{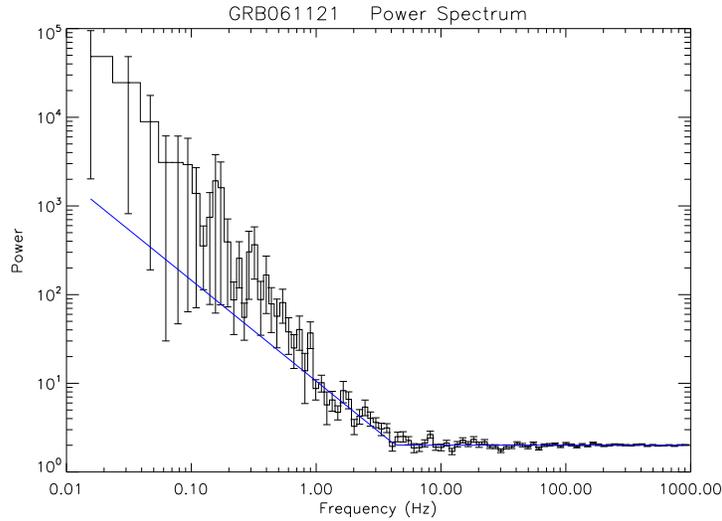}
  \caption{The power as a function of frequency. A broken power-law
  is fitted to the power spectrum to obtain the threshold
  frequency where the red noise (source signal) intersects the white noise
  (background).
  \label{power_spectrum}}
\end{figure}

The discrete Fourier transform of a series of N counts or numbers
is given by
\begin{equation}\label{eq:no1}
x_k = \frac{1}{N} \sum_{j=-N/2}^{N/2-1} a_j \, e^{-2 \pi \, i j
\,k/N},\,\,\,\,\,k=0,...,N-1;\,\,\,\,\,\ a_j = \sum_{k=0}^{N-1}
x_k \, e^{2 \pi \, i j
\,k/N},\,\,\,\,\,j=-\frac{N}{2},...,\frac{N}{2}-1
\end{equation}
To calculate the power spectrum, the Fast Fourier Transform (FFT)
algorithm was used~\citep{JenkinsWatta1968}. The power spectrum is
defined as \cite{Leahy1983}
\begin{equation}\label{eq:no2}
P_j \equiv \frac{2}{\sum
x_k}\,|a_j|^2,\,\,\,\,\,\,j=0,\,...,\frac{N}{2}.
\end{equation}
\\
We have used BAT event-by-event data and utilized the ``IDL
Extract''
software\footnote{http://idlastro.gsfc.nasa.gov/ftp/contrib/rxte/}
package to generate power spectra for our analysis. We divided the
burst into 2 segments and averaged the corresponding power spectra
to get the final spectra. A typical power spectrum from a GRB is
shown in Fig.~\ref{power_spectrum}. The low frequency power-law
component, sometimes referred to as ``red noise'', represents the
signal from the source. The high frequency region, called ``white
noise'', is background. The threshold frequency, $f_{\rm th}$, is
the frequency at which the red noise crosses the white noise
level. In order to extract the threshold frequency, the power
spectra were fitted by a broken power law
\begin{displaymath} 
P = \left\{ \begin{array}{ll} A \big(f/f_{\rm
th}\big)^{-\alpha},\,\,\,\,\,\,& f < f_{\rm
th},\\
A \big(f/f_{\rm th}\big)^{-\beta},\,\,\,\,\,\, & \rm otherwise.
\end{array} \right.
\end{displaymath}
\\
When calculating the isotropic peak luminosity ($L_{\rm iso}$), we
need to take into account the fact that in the rest frame of the
GRB, photon energies are higher than those in the observer frame.
The observed peak flux for the source--frame energy range
$1.0\,\rm{keV}$ to $10,000\,\rm{keV}$ was calculated using
observed spectral-fit parameters for the GRB sample and
appropriately z-corrected energy limits in the integral. The
isotropic peak luminosity is related to the calculated flux as
follows:
\begin{equation}\label{eq:no4}
L_{\rm iso}= 4 \pi d_L^{\,2} \, F_{\rm{obs}}.
\end{equation}
Here $F_{\rm{obs}}$ is the observed peak flux and the luminosity
distance, $d_L$, is given by,
\begin{equation}\label{eq:no5}
d_L=\frac{(1+z)c}{H_0}\int_{0}^{z} \frac{dz}{\sqrt{\Omega_M
(1+z)^3 + \Omega_k (1+z)^2 + \Omega_L}}.
\end{equation}
For the current universe we have assumed, $\Omega_M = 0.27$,
$\Omega_L = 0.73$, $\Omega_k = 0.0$ and a Hubble constant ($H_0$)
of $70\,\rm \, km \, s^{-1}\,Mpc^{-1} \,=\,2.268 \times
10^{-18}\,\rm s^{-1}$. The redshift measurements were taken from
online archives of the Gamma-Ray Burst Online Index
(GRBOX\footnote{http://lyra.berkeley.edu/grbox/grbox.php}) and
verified by using the GCN circulars.\footnote{Gamma-ray burst
Coordination Network (http://gcn.gsfc.nasa.gov)}

\section{Results}

\begin{figure}[h]
\includegraphics[angle=0, width=0.5 \textwidth]{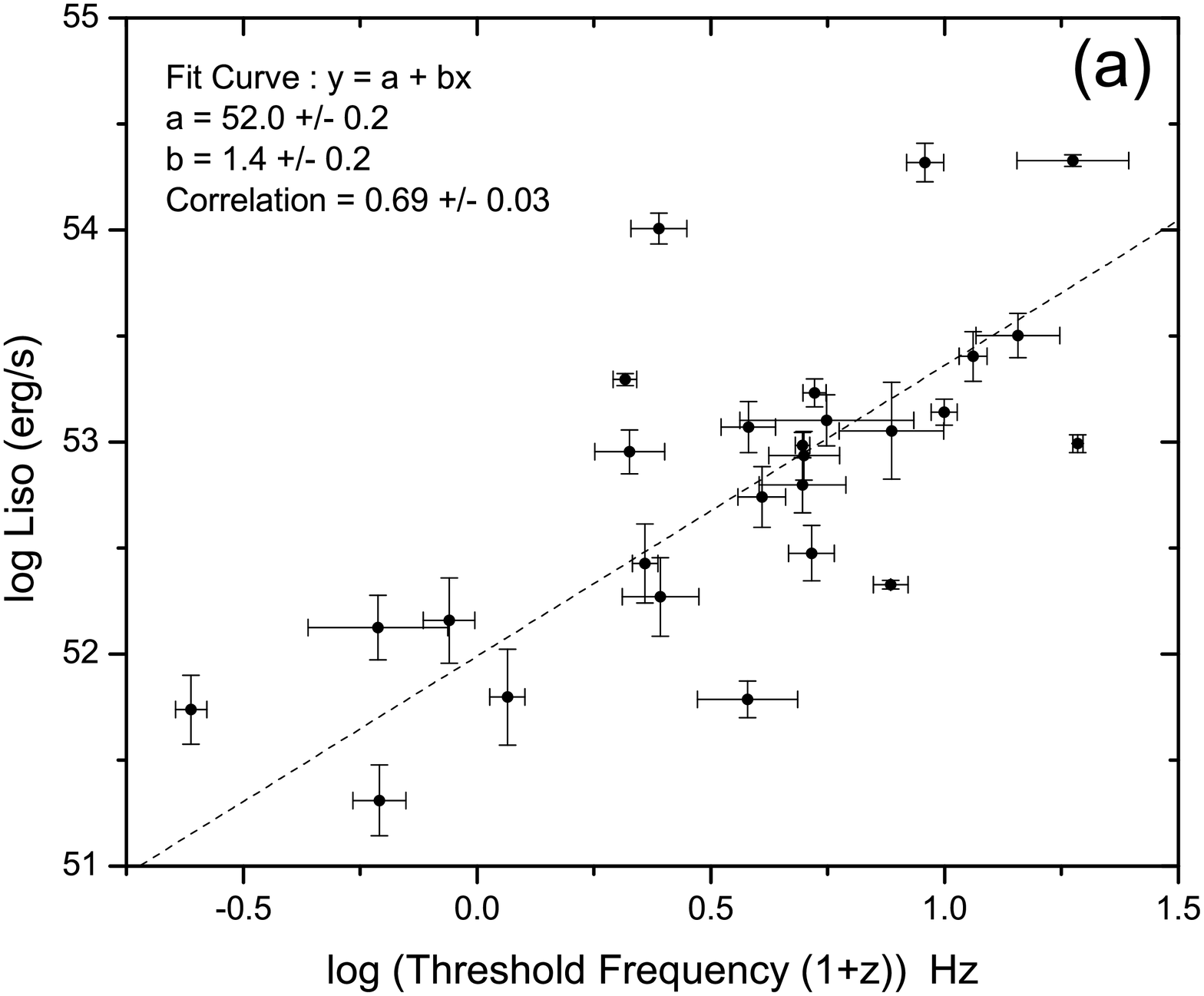}
\includegraphics[angle=0, width=0.5 \textwidth]{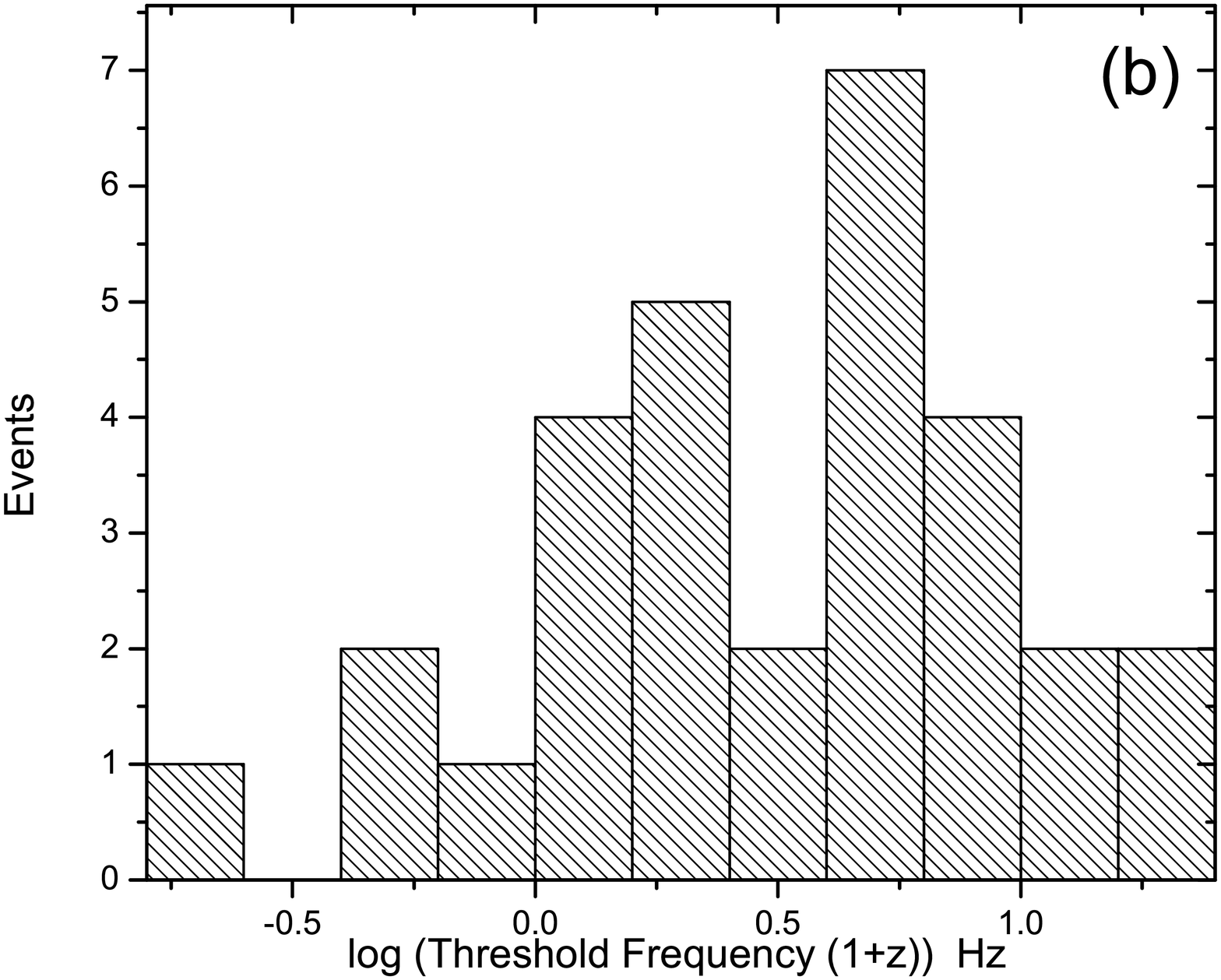}
  \caption{Panel (a): Isotropic peak luminosity as a function of threshold frequency. The correlation coefficient of the relation
  is $0.69 \pm 0.03$ with a chance probability of $3.4 \times 10^{-4}$.
  Panel (b): Distribution of the time-dilation-corrected threshold frequencies.
  \label{logLisovslogThFreq}}
\end{figure}

From a sample of 100 $Swift$ BAT GRBs with spectroscopically
confirmed redshifts, we selected 27 GRBs with good spectral
measurements. We analyzed event-by-event data of this sample and
obtained power spectra and fitted these with a broken power law
and extracted the threshold frequencies. We find the average
$\alpha$ for the sample is $1.12 \pm 0.05$ and $\beta$ is
consistent with zero for all fits. In
Fig.~\ref{logLisovslogThFreq} panel (a), we show the calculated
isotropic peak luminosity as a function of the extracted threshold
frequency (with the appropriate z correction). As seen in the
figure, the threshold frequency appears to be correlated with the
isotropic peak luminosity. The Pearson's correlation coefficient
is $0.69 \pm 0.03$, where the uncertainty was obtained through a
Monte Carlo simulation. The probability that the above correlation
occurs due to random chance is $3.4 \times 10^{-4}$. Our best-fit
yields the following relation between $L_{\rm iso}$ and $f_{\rm
th}$:
%
%
\\
\begin{equation}\label{eq:bestfit}
\log L_{\rm iso} = (52.0 \pm 0.2) + (1.4 \pm 0.2) \log (f_{\rm
th}(z+1)).
\end{equation}
\\
Panel (b) in Fig.~\ref{logLisovslogThFreq} shows a histogram of
the threshold frequencies. The lowest redshift-corrected threshold
frequency in the sample is approximately 0.2 Hz and the largest is
about 20 Hz. Our results imply the smallest variability time scale
to be approximately 50 msec.
\\ \\

\section{Discussion}

\begin{figure}[htp]
\includegraphics[angle=0, width=0.5 \textwidth]{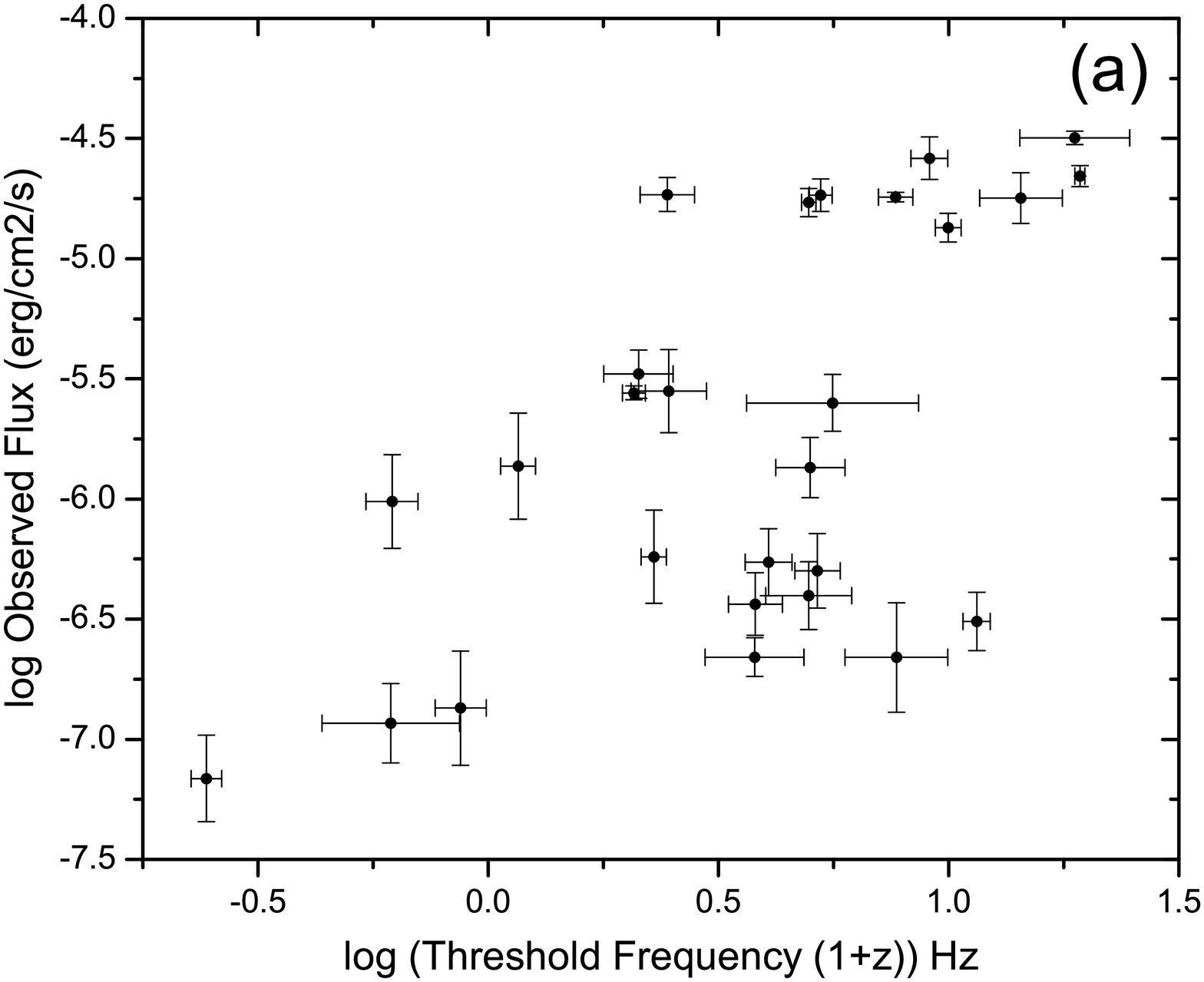}
\includegraphics[angle=0, width=0.48 \textwidth]{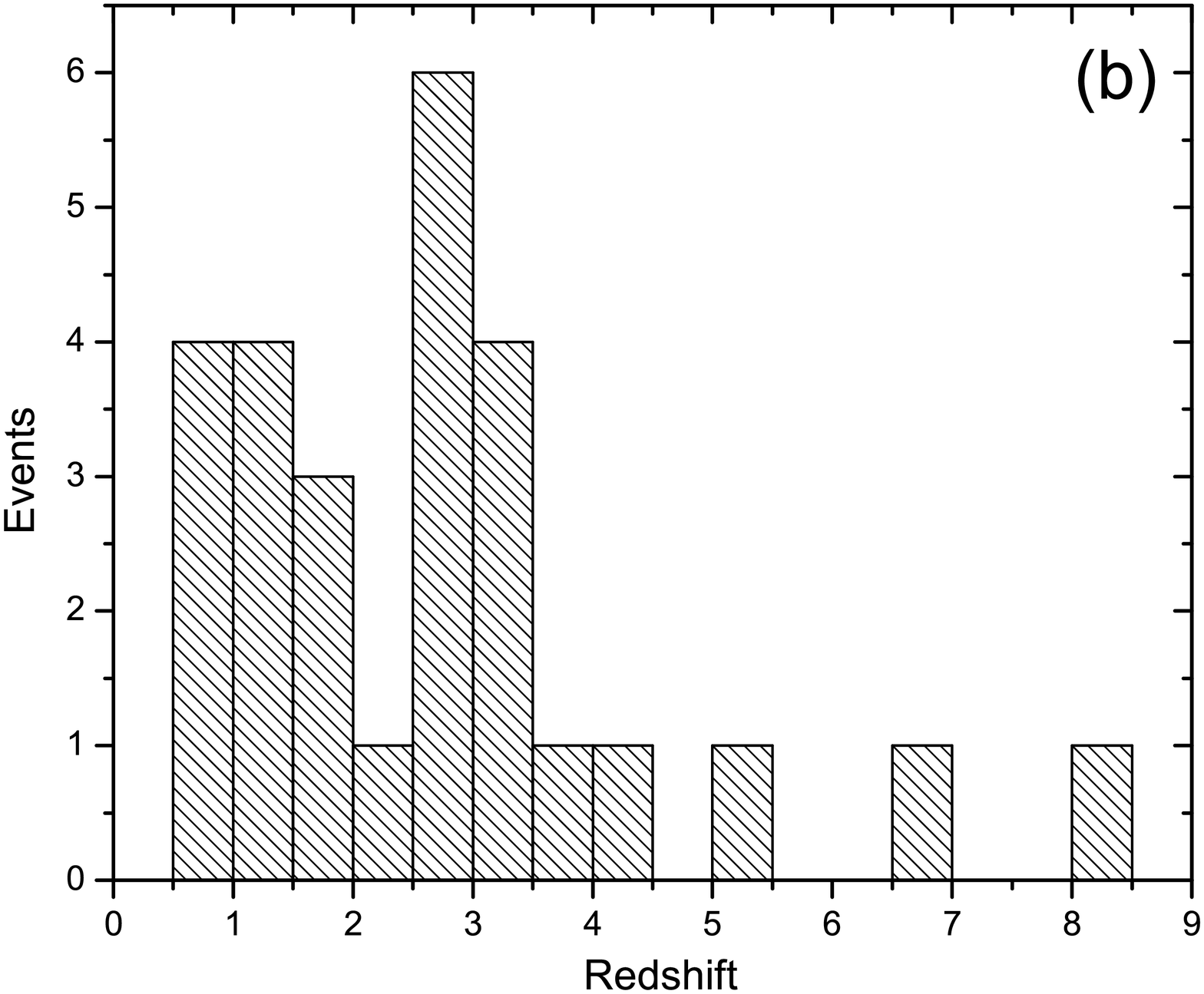}
  \caption{Panel (a): Isotropic peak luminosity as a function of observed flux.
  Panel (b): The histogram of redshifts in the sample.
  \label{zHistogram}}
\end{figure}

It is important to investigate how the observed brightness of GRBs
affects the extracted threshold frequencies because the threshold
frequency also appears to be correlated with the observed
brightness (see Fig.~\ref{zHistogram} panel (a)) and this
potential observational bias needs further investigation.
\\ \\
FFT power is roughly proportional to the brightness of the burst.
Hence when a burst of a given luminosity is closer and therefore
brighter, the red noise is enhanced above the constant Poisson
noise level (white noise) and results in a larger threshold
frequency. Given a burst of luminosity $L_{\rm iso}$ and frequency
$f_{\rm th}$, the same burst with a different luminosity $L_{\rm
iso}'$ would fall on the following $L_{\rm iso}' - f_{\rm th}'$
line:
\begin{equation}\label{eq:no7}
\log L_{\rm iso}' = \log (L_{\rm iso}/f_{\rm th}^\alpha) + \alpha
\log f_{\rm th}'.
\end{equation}
The mean value of $\alpha$ for our sample is $1.12 \pm 0.05$,
which does not fully explain the slope of $1.4 \pm 0.2$ (see
Fig.~\ref{logLisovslogThFreq}) obtained from the $L_{\rm
iso}$-$f_{\rm th}$ correlation.
\\ \\
We note also that the uneven distribution of redshifts (see
Fig~\ref{zHistogram} panel (b)) might play a role in the
correlation with flux. Roughly half of the sample lies in the
redshift range 1.5 to 3.5 where the luminosity distance changes
only by factor of 2.8, hence, the correlation between $L_{\rm
iso}$ and $f_{\rm th}$ partially trickles down to the flux as
well.
\\
\\
In conclusion, we argue that the FFT threshold frequency is a
parameter that provides a suggestive measure of gamma-ray bursts
variability. For the sample of 27 GRBs analyzed, our results imply
the smallest variability time scale is approximately 50
milliseconds. The apparent correlation between the isotropic peak
luminosity and the threshold frequency needs further investigation
for observational bias. If confirmed, this correlation is
potentially useful as a probe of GRB microphysics. It may also
prove useful as a redshift estimator.

\end{document}